\documentclass[12pt,a4paper]{article}
\usepackage{jheppub}

\usepackage{epsfig,multirow} 
\usepackage{amscd}
\usepackage[matrix,arrow,curve]{xy}
\usepackage{verbatim}
\usepackage{latexsym}
\usepackage{amsfonts,amsthm,amsmath}

\renewcommand{\a}{\alpha}

\renewcommand{\l}{\lambda}

\subheader{UCSD-PTH-11-15}
\title{\begin{center}
$\mathcal Z $ Extremization in Chiral-Like \\ Chern Simons  Theories 
\end{center}}
\author[a]{Antonio Amariti,}
\author[b]{Massimo Siani}

\affiliation[a]{Department of Physics, University of California, \\
San Diego La Jolla, CA 92093-0354, USA}
\affiliation[b]{Instituut voor Theoretische Fysica, Katholieke Universiteit Leuven,\\
Celestijnenlaan 200D, B-3001 Leuven, Belgium}

\emailAdd{antonio.amariti@physics.ucsd.edu}
\emailAdd{massimo.siani@fys.kuleuven.be}

\abstract{We study the localized free energy on $S^3$ 
of three-dimensional $\mathcal{N}=2$ Chern-Simons 
matter theories at weak coupling. 
We compute the two loop $R$ charge in three different ways, namely by the standard
perturbative approach, by extremizing the localized partition function at finite $N$ and by
applying the standard saddle point approximation for large $N$.
We show that the latter approach does not 
reproduce the expected result
when chiral theories are
considered. We circumvent these problems by restoring
a reflection symmetry  on the eigenvalues in the free energy. Thanks to this symmetrization
we find that the three methods employed agree.
In particular we match the computation for a model whose  four dimensional parent 
is the quiver gauge theory describing  D$3$ branes probing the Hirzebruch  surface.
We conclude by commenting on the application of our results
and to the strong coupling regime.
}

\begin{document}
\maketitle

\section{Introduction}

Localization of  three dimensional supersymmetric field theories on $S^3$ 
has recently attracted many  investigations.
Indeed it has been observed that it 
is not only an academic exercise, but this procedure 
offers a simple way to extract many information about  SCFTs.
The partition function and then the free energy computed by this procedure 
reduce to a matrix model which contains some
quantum information once  only
the one loop determinants are evaluated.
This technique, first applied to ${\cal N} \geq 3$
superconformal Chern Simons matter theories  
in \cite{Kapustin:2009kz}, made it possible the computation of the 
large $N$ scaling of the free energy in $\mathcal{N} \geq 3$ 
theories \cite{Drukker:2010nc,Herzog:2010hf}, which are supposed to describe
M$2$ branes probing a conical singularity which basis is a CY$_4$ \cite{Aharony:2008ug}.
This is a non trivial check of the conjectured duality because it was observed that the
$N^{3/2}$ scaling of the free energy obtained from the supergravity side is recovered even by
a direct computation in a strongly coupled field theory.

The ${\cal N} \geq 3$ partition function trivially depends on the $R$ charge of the matter fields,
because the supersymmetry algebra implies that they keep their canonical dimensions also at the
quantum level.
A more complicated situation appears in the $\mathcal{N}=2$ case where the $R$ symmetry 
is abelian and it can mix with the other abelian symmetries of the theory.
In this case the free energy is a function of
the $R$ charge, and the properties of the $\mathcal{N} = 2$ theories 
require the knowledge of the exact $R$ charge current appearing in the superconformal algebra.
Surprisingly, as first observed in \cite{Jafferis:2010un},  the exact $R$ charges have the special property
of extremizing the free energy $\mathcal{F}$.\footnote{This result was derived from a \emph{mysterious} holomorphy, which 
origin was then explained in \cite{Festuccia:2011ws}.}
In every known example this extremization is a maximization, and identifying the free energy  as a good candidate for
counting the number of degrees of freedom
led to conjecture a three-dimensional c-theorem, known as ${\cal F}$-theorem
 \cite{Jafferis:2011zi,Klebanov:2011gs,Gulotta:2011si}.
In \cite{Klebanov:2011gs,Amariti:2011xp}, the $\mathcal{F}$-maximization was
found to be valid along the whole RG flows of a large class of weakly coupled theories, even without supersymmetry
\cite{Klebanov:2011gs}, corroborating the validity of an $\mathcal{F}$-theorem.
Other properties like the 
relation with the supergravity computations and the volumes of 
the $AdS$ dual geometry  \cite{Martelli:2011qj,Cheon:2011vi,Jafferis:2011zi,Fuji:2011km} 
and the relation between the Lagrange multipliers enforcing the marginality of the superpotential terms and
the coupling constants along the RG flows  \cite{Amariti:2011xp} were then observed.

In many classes of theories the $\mathcal{F}$ maximization procedure has been 
successfully applied and the  known perturbative and non perturbative
results 
\cite{Martelli:2011qj,Cheon:2011vi,Jafferis:2011zi,Amariti:2011hw,Niarchos:2011sn,Minwalla,Amariti:2011da,Morita:2011cs} 
have been recovered.  
Anyway there is a whole class of theories
where the procedure has not been completely understood yet.
This class is typical of theories with low supersymmetry ($\mathcal{N}=1$ in 4d and $\mathcal{N}=2$ in
3d) and consists of theories with a chiral like field content. In the language of quiver gauge theories
they are represented as 
nodes (groups) connected by oriented edges (bi-fundamentals and adjoint fields), such that 
there are edges where the number of oriented arrows in one direction differs 
from the number of oriented arrows in the opposite direction.
These theories are called chiral gauge theories.
It was  observed that at large $N$ and strong coupling ($N \gg k$) the free energy does not scale as $N^{3/2}$ 
as predicted by the conjectured supergravity dual and the $\mathcal{F}$ extremization procedure does not reproduce the 
volume computation.
Understanding this mismatch is an open problem. Indeed if the large $N$ scaling 
is different from the one expected in supergravity these theories cannot describe the motion of M$2$ branes probing a
toric CY$_4$. On the other hand if these theories provide the correct dual SCFT, then there must
be subtleties in the large $N$ computation of the free energy.

The discussion above motivates the study of the free energy of theories with a chiral like field content.
In this paper we discuss the large $N$ scaling limit of $\cal{F}$, but  in the weakly coupled regime, 
where the CS level $k$ is larger than the gauge group rank $N$, and the 't Hooft coupling $N/k$ is small.
\footnote{We define the 't Hooft coupling to be $N/k$ even though $N$ will be sometime kept finite in the large $k$ limit.}
In this regime, the $R$ charges of the matter fields can be computed in several  independent ways. We are interested
in comparing the $\cal F$-extremization technique for large $k$ and finite $N$ as well as its large $k$, large
$N$ saddle point approximation to the standard perturbative evaluation. We show that in the case of vector-like
theories the three approaches give exactly the same result. 
Then we switch to  a chiral theory
with one gauge group with $N_f$ fundamental and $N_{\widetilde f}$
anti-fundamental with $N_f \neq N_{\widetilde f}$. In the latter theory we first compute 
the partition function for finite $N$ observing that the corresponding $R$ charge agrees with the
expected one. Then we show that the naive saddle point equations at large $N$ do not reproduce the expected
large $N$ result.
We observe that, at this order, this is related to the explicit breaking of the
reflection symmetry, on the large $N$ saddle point equations,
acting on the eigenvalues of the Cartan subgroup of the gauge group. By restoring this
symmetry, the $\mathcal{F}$-extremization procedure in its 
large $N$ saddle point approximation gives the same answer as the large
$N$ limit obtained from the other two computations.
We also check our proposal in the case of multiple gauge groups at the perturbative level by looking at a quiver gauge theory
with the same field content and superpotential of the four dimensional $\mathbb{F}_0$ model but
with a CS term  for each gauge group and large levels $\{k_i\}$ (we refer to this theory as 
$\widetilde{\mathbb{F}_0}_{\{k_i\}}$). This theory 
in the strongly coupled regime is conjectured to describe M$2$ branes probing a CY$_4$ geometry 
(whose geometrical details depend from the choice of the CS levels).
Moreover we observe that the Lagrange multiplier proposal of \cite{Amariti:2011xp} can be extended to this chiral example 
straightforwardly. We finally consider a generalization of this theory to an arbitrary number of bi-fundamental fields
and show that our method still gives the expected result.

The paper is organized as follows. In section \ref{sec1} we review the basic aspects of the localized partition function and its 
relation with the exact $R$-charge. In section \ref{sec2} we show the computation of the free energy at large $N$ but small 't Hooft coupling. 
We discuss both vector like and chiral like examples, and observe that the latter cases match with the
perturbative computation only after we restore the reflection symmetry on the Cartan subgroup.
In section \ref{sec3} 
we verify the Lagrange multiplier conjecture in the case of $\widetilde{\mathbb{F}_0}_{\{k_i\}}$.
Then we conclude, and comment on some possible application of our method in the strongly coupled regime.

\section{$\cal F$ extremization} \label{sec1}

As mentioned in the introduction, the localization technique has been shown to be
a powerful tool to extract physical quantities in three-dimensional field theories.
In the case of $\mathcal{N}\geq 3$ theories \cite{Kapustin:2009kz} the $R$ symmetry
is non abelian and the supersymmetry algebra   
constraints the matter fields to preserve their classical scaling dimension.
In the $\mathcal N = 2$ case, the classical value of the scaling dimension of the matter fields
is not generically preserved along the RG flow.
Indeed in this case the  $R$-symmetry group
$SO(2)_R\simeq U(1)_R$ is abelian and it can mix with all the other abelian flavor symmetries
of the theory. Moreover it can also mix with topological symmetries related to the diagonal monopoles.\footnote{Anyway we can neglect this contribution at the lower orders at perturbative level. In the Appendix \ref{appmax} we show that even if the monopole contributions are
added they do not affect the two loop results.}
The
full localized partition function on $S^3$ is \cite{Jafferis:2010un,Hama:2010av}
(up to an overall factor which is irrelevant for our purposes)
\begin{equation}
\mathcal{Z}_{S^3} = \int  \text{d}[u]e^{ i \pi k \text{Tr}u^2 }
{\det}_{\text{adj}} \sinh (\pi u ) \prod_{\Phi} {\det}_{\rho_{R_\Phi}} e^{l\left( 1-\Delta_{R_\Phi}+i \rho_{R_\Phi}(u) \right)} 
\label{eq:partfunc}
\end{equation}
The different contributions to this formula are
\begin{itemize}
\item The measure d$[u]$ is the measure over the Cartan of the gauge group. For example for a $U(N)$ model
it is 
\begin{equation}
\text{d}[u]=\prod_{i=1}^{N} \text{d}u_i
\end{equation}
\item
The exponential $e^{ i \pi k \text{Tr}u^2 }$ is the contribution coming from a CS term with level $k$.
The trace is over the fundamental, namely Tr$u^2=u_1^2+\dots+u_N^2$ for the $U(N)$ case.
The YM contribution vanishes because $g_{YM}$ is dimensionful in three dimensions.
\item
The determinant over the adjoint is the product of the roots and it comes from the 
one loop determinant of the vector field. In the $U(N)$ case it is explicitly
\begin{equation}
{\det}_{adj} \sinh \pi u=\prod_{i<j} \sinh^2(\pi(u_i-u_j))
\end{equation}
\item
The last contribution is the one loop determinant of the matter fields. For every field $\Phi$ in the representation $R_\Phi$
of the gauge group the determinant is computed over the weights $\rho_{R_\Phi}$ of the representation.
The function $l(z)$ has been found in \cite{Jafferis:2010un,Hama:2010av} after the zeta regularization of the otherwise 
divergent determinant. It is explicitly
\begin{equation}
l(z) = -\frac{i \pi}{12} -z \, \text{Log} \left( 1-e^{2 i \pi z}\right) +\frac{1}{2} i\left(\pi z^2+\frac{\text{Li}_2\left(e^{2 \pi i z}\right)}{\pi}\right)
\label{eq:lz}
\end{equation}
The partition function has an explicit dependence on the $R$ charge $\Delta$ through this function, which
then represents the main difference with respect to the more supersymmetric models. Indeed when ${\cal N} \geq 3$
the non abelian nature of the $R$-symmetry group constraints $\Delta$ to  acquire the classical value $1/2$
at the quantum level (no mixing is possible in this case) and the formula obtained in \cite{Kapustin:2009kz} is recovered.
\end{itemize}

The proposal made in \cite{Jafferis:2010un} consists of extracting the exact $R$ charge by extremizing the partition
function (\ref{eq:partfunc}) with respect to $\Delta$.
Many examples were worked out both at
strong and at weak coupling,
comparing the results with the AdS/CFT predictions and the perturbative evaluations, respectively. In all the cases,
the $\mathcal{Z}$-extremization reproduced the expected results.

A general strategy for computing the integral (\ref{eq:partfunc}) is still lacking, but approximate computations 
of the free energy have been performed. For example 
at large $k$ \cite{Amariti:2011hw,Amariti:2011da}
the saddle point
contribution to the integral comes from $u_i\sim0$, where the integrals reduce to simple gaussian integrals and the perturbative results are computed in terms of the small 't Hooft coupling. Moreover in \cite{Minwalla} the computation was simplified in the case of $k\gg N\gg1$.
In this case the saddle point equations can be solved perturbatively order by order in the eigenvalues. At the lowest order they  obey a Wigner distribution and the classical $R$ charge is $R=1/2$. The solutions to the higher order equations
are associated to the quantum corrections in the QFT
perturbative expansion.

The strong coupling regime has been deeply investigated in
\cite{Martelli:2011qj,Jafferis:2011zi}.\footnote{
A different approach for the $\mathcal{N}\geq 3$ case has recently appeared in \cite{Suyama:2011yz}.}
The saddle point equations have been solved only for  particular classes of theories, like the necklace 
non-chiral quiver gauge theories.\footnote{Another class of theories in which the $N^{3/2}$
 scaling behavior and the equivalence with the volume minimization was found consists of 
 non chiral quiver gauge theories with the addition of chiral flavors \cite{Benini:2009qs,Jafferis:2009th}.}
These theories correspond to quiver gauge theories with bi-fundamentals connecting 
two adjacent groups, and a vector like field content. Indeed in this case it was shown that the 
eigenvalues at large $N$ scale as $u_i= \sqrt N x_i + i  y_i$. This scaling and the structure of the quiver cancel the
long range interactions among the eigenvalues and leave a local structure for the free energy.
The $N^{3/2}$ scaling at large $N$ has been matched with the supergravity calculation and with the volume 
minimization. The validity of the $\mathcal{F}$-maximization has not been confirmed yet
 for the theories with a chiral like field content. 
The main obstruction is that in these cases the long range interactions among the eigenvalues cannot
be cancelled as in the non-chiral cases, and the scaling of the free energy  usually differs from
 the one expected by the supergravity computation.

Many interpretations are possible for this mismatch. It is possible that  these theories do not describe
the motion of M$2$ branes in the CY$_4$ geometry and cannot match with the
AdS/CFT predictions.  Another possibility is that there is some problem in the large $N$ approach.
Motivated by this mismatch, in the rest of the paper we study the large
$N$ saddle point equations for chiral like theories in the
perturbative regime $k\gg N\gg1$. We will show that a well defined large $N$ limit in this regime
requires some care.

Weakly coupled theories have the advantage that we have many tools to carry out the computation, providing us with
reliable checks about the validity of the results. Indeed, we will perform our computation in three different ways. First, we
present the standard perturbative evaluation of Feynman diagrams. Secondly, we show that when we consider the finite $N$
partition function in the perturbative regime, we obtain the same results as the Feynman diagram evaluation. Finally, we
show that in the large $N$ limit, namely in the saddle point approximation, some assumptions are breaking down, and we
present a way to extract the right result. We show that our procedure works in several examples. Based on these results,
we comment on the large $N$ limit in the strongly coupled regime.

\section{Perturbative regime at large $k$ and $N$} \label{sec2}

Before we start our analysis a comment is in order.
Even though in three dimensions there are no local gauge anomalies,  the gauge
invariance may require the introduction of a classical CS term which breaks parity 
\cite{Redlich:1983dv,AlvarezGaume:1983ig}.
This is usually referred to as parity anomaly. For example in the abelian case
with multiple $U(1)$'s there is a parity anomaly if
\begin{equation} \label{sumpar}
\frac{1}{2} \sum_{\text{fermions}} (q_f)_i(q_f)_j \in \mathbb{Z}+\frac{1}{2} 
\end{equation}
where $(q_f)_i$ is the charge of the fermion $f$ under the $U(1)_i$. 
If it is the case a semi integer CS term must be added to restore parity.\footnote{Anyway in the perturbative case $1/k$ is a continuous variable, and even if parity is broken and 
the level is shifted by $1/2$ the computation is still valid.}
In the rest of the paper we restrict  to the cases in which (\ref{sumpar}) is
integer.

To the best of our knowledge the perturbative computations of the $R$ charges in the chiral theories the
we are considering have never appeared in the literature. Anyway we will only present the
final result and stress that it agrees with the other methods employed. The interested reader is referred
to \cite{Avdeev:1992jt} for the details of the standard perturbative approach.

In the rest of this section we will only consider $SU(N)$ models where it is
possible to associate an $R$ charge to the fundamentals or
bi-fundamentals. Anyway at large $N$ the difference between $U(N)$ and $SU(N)$ 
is sub-leading and we can trust
in the extremization of  the free energy associated to the $U(N)$ 
model. Indeed one can observe that at the order we are interested in
the eigenvalues sum up to zero even in the $U(N)$ case,
which corresponds to the $SU(N)$ traceless condition.
In the Appendix \ref{appmax} we explicitly observe the agreement 
at large $N$ of the $U(N)$ and $SU(N)$ computations.

\subsection{$SU(N)_k$ with $N_f$ fundamentals and anti-fundamentals} \label{secFFB}

We now apply the saddle point approximation to the weak coupling regime of a vector-like Chern-Simons theory.
We couple the ${\cal N}=2$ vector supermultiplet $V$ to $N_f$(=$N_{\widetilde f}$) pairs $(\phi,\tilde\phi)$ of fields in the (anti)fundamental
representation of the gauge group $G=SU(N)_k$. For simplicity we consider a vanishing superpotential, but the extension
to a more general case is straightforward.

The three-dimensional localized partition function for this model is 
\begin{eqnarray} \label{ptfaf}
\mathcal{Z} &=& \int \prod_{i=1}^{N} d u_i \text{Exp}\left(N^2\left( \frac{i \pi}{\lambda N}\sum_{i=1}^{N}u_i^2 +\frac{1}{N^2} \sum_{i\neq j} \log \sinh \left( \pi u_{ij} \right)+\right. \right. \nonumber \\ &&~~~~~~~~~~~~~~~~~~~~~~~~~~~~~~~~~~~~~~~~~+\left. \left.
\frac{N_f}{N^2} \sum_{i;\eta=\pm 1} l(1-\Delta+i \eta u_{i})\right)\right)
\end{eqnarray}
where we defined the 't Hooft coupling $\l\equiv N/k$.
For large enough $N$, the main contribution to $\cal Z$ comes from the extremum of the argument of the exponential.
In order to find the eigenvalues $\{u\}$  which correspond to this minimum, we write the following saddle point equations
\begin{eqnarray} \label{saddle2}
\frac{ i}{\lambda} u_i + \frac{1}{N} \sum_{j\neq i} \coth(\pi u_{ij}) -\frac{i N_f}{2 N}\sum_{\eta=\pm 1} \eta (1-\Delta+i \eta u_{i} )  \cot (1-\Delta+i\eta u_{i} ) = 0
\end{eqnarray}
and we substitute the corresponding solution into the extremization equation for the $R$ charge
\begin{equation} \label{extr2}
\text{Re} \left( \sum_{i;\eta=\pm 1}(1-\Delta+i \eta u_{i})\cot(\pi(1-\Delta+i \eta u_{i}))\right)=0
\end{equation}
In the perturbative regime $\l \ll 1$ we expand the eigenvalues and the $R$ charge as \cite{Minwalla}
\begin{equation}
\begin{split}
u_i &= \sum_{n=0}^{\infty} u_i^{(n)} \l^{\frac{1}{2}+n} =\sqrt{\l} \left( u_i^{(0)} + \l u_i^{(1)} + \ldots \right) \\
\Delta &= \Delta^{(0)} + \l \Delta^{(1)} + \l^2 \Delta^{(2)} + \ldots
\label{charge2}
\end{split}
\end{equation}

To lowest order, equation (\ref{extr2}) sets $\Delta^{(0)}=1/2$, which is the classical scaling dimension, as expected
 in the perturbative case. By substitution in (\ref{saddle2}), we find that the $u_i^{(0)}$ satisfy
\begin{equation}
i u_i^{(0)} +\frac{1}{\pi N} \sum_{j(\neq i )} \frac{1}{u_i^{(0)}-u_{j}^{(0)}} = 0
\label{eq:orderm12}
\end{equation}
whose solution is known in the large $N$ limit: we rotate the eigenvalues according to $u_i^{(0)} = y_i \exp{(i \pi/4)}$
and because the variables $y_i$ become dense we substitute them with the continuos variable $y$. The eigenvalue
distribution $\rho(y)$ has support on the interval $(-\sqrt \frac{2}{\pi},\sqrt \frac{2}{\pi})$ and takes the value
\begin{equation}
\rho(y) =\sum_{i=1}^{N} \delta(y-y_i) = \sqrt{\frac{2}{\pi}-y^2}
\end{equation}

Till now, we recovered the classical behavior of the field theory. The quantum corrections are contained in the higher
order expansions of (\ref{saddle2}) and (\ref{extr2}). In particular, one finds that the distribution $\rho(y)$ is not
changed and as a consequence $\Delta^{(1)}=0$, again in agreement with perturbation theory.

We substitute back these results into (\ref{saddle2}) and expand till the next nontrivial order.
After some manipulations we get the equation
\begin{equation} \label{corrNf}
\frac{1}{\pi N} \sum_{j \neq i} \frac{u_j^{(1)}}{u_i^{(0)}-u_j^{(0)}}=\frac{\pi}{6 }\left(\frac{3 N_f}{N}-2\right) u_i^{(0)}
\end{equation}
We again rotate the variables in the complex plane as
$u_i^{(0)} \rightarrow y_i = e^{-\frac{i \pi}{4}} u_i^{(0)}$ and 
$u_i^{(1)} \rightarrow v_i(y_i) = e^{-\frac{i \pi}{4}} u_i^{(1)}$, and substitute the discrete $y_i$ and $v_i$ with the
continuos variables $y$ and $v(y)$. Using the same technique explained in \cite{Minwalla} we solve the corresponding
equation obtaining
\begin{equation}
v(y) = -\frac{i \pi}{12} \left( \frac{3 N_f}{N} -2 \right) y
\end{equation}

We now have all the necessary ingredients to compute the two-loop $R$ charge. Indeed, one first notes that at
 $\mathcal{O}(\lambda^{3/2})$ equation (\ref{extr2}) is an identity, then one finds a nontrivial equation at $\mathcal{O}(\lambda^2)$. Again, the latter can be solved by passing to the
continuum limit which results in
\begin{equation}
\Delta = \frac{1}{2} - \l^2 \frac{N_f}{2 N}
\label{eq:sameNf}
\end{equation}
Thus, we obtained the large $N$ limit of the full two-loop perturbative result \cite{Gaiotto:2007qi} 
already computed for finite $N$ in \cite{Amariti:2011hw}.

\subsection{$SU(N)_k$ with different number of fundamentals and anti-fundamentals}

\subsubsection{Extremizing the finite $N$ partition function}
\label{sec:FullZ}

For simplicity, we consider a ${\cal N}=2$ model with gauge group $SU(N)$ and
$N_f$ massless chiral fields $\phi^i_a$, $a=1,\ldots,N$, $i=1,\ldots,N_f$ in the fundamental representation of the gauge group.
The resulting partition function reads
\begin{equation} \label{parf}
{\cal Z}= \int \prod_{i=1}^{N} d u_i \text{Exp} \left(\!i \pi k \sum_{i=1}^{N}u_i^2 + \sum_{i\neq j} \log \sinh \left( \pi u_{ij}\right)
\!+\! \sum_{i} N_f l(1\!-\!\Delta\!+\!i  u_{i})\!\! \right) \delta\!\left(\sum_{i=1}^N u_i\!\right)
\end{equation}
By using the strategy  of \cite{Marino:2002fk,Aganagic:2002wv}
 in the large $k$ limit, keeping $N$ fixed, (\ref{parf}) reduces to (up to an overall factor)
\begin{equation}
\begin{split}
{\cal Z} &\simeq 1152 k^2 N (4+a^2 N_f N \pi^2)-(N^2-1)\pi^2 (4 (16 N^5+9 (1-4 a)^2 N_f^2 N (N^2+1)+ \\
&  +12 N_f (3+2 N^4-8 a (3-N^2+N^4)))-a^2 N_f 
(N (9 N N_f^2 (N^2+1)+\\&+12 N_f (3+6 N^2-2 N^4)+16 N (18-6 N^2+N^4))-576) \pi^2 ) +\\
&+ 48 i k N (N^2-1) \pi  (16 N-N_f (12-a (48+a (4 (N^2-3)-3 N_f N) \pi^2)))
\end{split}
\end{equation}
where $\Delta=1/2+a$. Upon extremization, we obtain the $R$ charge of the fields
\begin{equation}
\Delta= \frac{1}{2} - \frac{(N N_f - 2) \left(N^2 -1 \right)}{4 N^2 k^2}
\label{eq:equalNf}
\end{equation}
in full agreement with the perturbative result. We also performed a similar computation to extract the $R$ charge
of a $SU(N)$ gauge theory coupled to $N_f$ fundamental fields $\phi_a^i$, $i=1,\ldots,N_f$
and $N_{\widetilde f}$ antifundamental fields $\tilde \phi_{i^\prime}^a$, $i^\prime=1,\ldots,N_{\widetilde f}$, with no
superpotential term. The result, which again agrees with the perturbative one, is
\begin{equation}
\Delta_1 = \Delta_2 = \frac{1}{2} - \frac{\left(N \left(N_f+N_{\widetilde f} \right) - 2\right) \left(N^2 -1 \right)}{4 N^2 k^2}
\label{eq:differentNf}
\end{equation}
where $\Delta_1$ ($\Delta_2$) is the $R$ charge of $\phi$ ($\tilde \phi$).\footnote{The results (\ref{eq:equalNf})
and (\ref{eq:differentNf}) can be
understood by noticing that at the two-loop order there is no difference for the gauge contribution between a field
and one in the complex conjugate representation. What matters is the number of fields. This explains why $N_f$ in
(\ref{eq:equalNf}) is replaced with $N_f+N_{\widetilde f}$ in (\ref{eq:differentNf}). This argument also extends previous
weak coupling computations in vector-like theories without ${\cal N}=3$ deformations \cite{Avdeev:1992jt,Amariti:2011xp,Amariti:2011hw,Amariti:2011da,Akerblom:2009gx,Bianchi:2009ja,Bianchi:2009rf} to chiral field theories.}

This shows that, at least at weak coupling, the extremization procedure
gives the correct exact $R$ charge.
While we have no full reliable check of the validity of this statement either at higher orders in
perturbation theory or at strong coupling, we do not see any obstruction for its validity.

\subsubsection{The saddle point approach}

We now apply to the $SU(N)$ gauge theory coupled to $N_f$ ($N_{\widetilde f}$) (anti)fundamental fields described at the
end of the previous subsection the saddle point approximation, along the lines described in section \ref{secFFB}.
When the number of fundamentals and the number of anti-fundamentals do not coincide the partition function (\ref{ptfaf})
becomes
\begin{eqnarray} \label{ptfaf1}
\mathcal{Z}& =& \int \prod_{i=1}^{N} d u_i \, \, \text{Exp}\left\{ N^2 \left( \frac{i \pi}{\lambda N}\sum_{i=1}^{N}u_i^2 +\frac{1}{N^2} \sum_{i\neq j} \log \sinh \left( \pi u_{ij}\right) \right. \right. \nonumber\\
&+&\left. \left. \frac{1}{N^2} \sum_{i} N_f l(1-\Delta_1+i  u_{i})+N_{\widetilde f} \, l(1-\Delta_2-i  u_{i}) \right) \right\}
\end{eqnarray}
As opposed to (\ref{ptfaf}), the integrand of (\ref{ptfaf1}) is not invariant under the transformation $u_i \rightarrow -u_i$.
While the latter "parity" transformation is an obvious symmetry of the integrand of the partition function of all the vector-like
three-dimensional models, such symmetry is explicitly broken in all the chiral models. As a result, we will now show that this
makes the ansatz (\ref{charge2}) inconsistent for the eigenvalues $\{u\}$.

Indeed, by solving order by order the saddle point  equation
\begin{eqnarray} \label{saddlefft}
\frac{ i}{\lambda} u_i + \frac{1}{N} \sum_{j\neq i} \coth(\pi u_{ij}) &-&\frac{i N_f}{2 N} (1-\Delta+i  u_{i} )  \cot (1-\Delta+i u_{i} )+
\nonumber \\ &-&\frac{i N_{\widetilde f}}{2 N} (1-\Delta-i  u_{i} )  \cot (1-\Delta-i u_{i} ) = 0
\end{eqnarray}
and the extremization equation,
one finds that the solutions are the same as in the vector like case, but we also get two
inconsistent equations. More precisely the saddle point equation (\ref{saddlefft}) at order 
$\mathcal{O} (\lambda)$ gives ${u_{i}^{(0)}}^2=0$
which contradicts the ${\cal O}(\l^{-1/2})$ equation (\ref{eq:orderm12}). This implies that some extra term has to be added in the
expansion of the eigenvalues, such that the equation at this order can be solved. 
For example if the expansion is modified as
\begin{eqnarray}
u_i = \sum_{n=0}^{\infty} u_i^{(n/2)} \lambda^{\frac{n+1}{2}} = \sqrt{\l} \left( u_i^{(0)} + \l^{1/2} u_i^{(1/2)} + \l u_i^{(1)} +
 \l^{3/2} u_i^{(3/2)} + \ldots \right)
\label{eq:charge3}
\end{eqnarray}
then both the saddle point and the extremization equations can be consistently solved order by order in $\l$ up to the
two-loop level. These require $u_i^{(1/2)}$ to vanish but $u_i^{(3/2)}$ not to vanish, while $u_i^{(0)}$ and $u_i^{(1)}$
obey equations analogous to those in section \ref{secFFB}.
Indeed in this case we  can solve the equation at order $\mathcal{O}(\lambda)$  for $u_i^{(3/2)}$
\begin{equation}
i u_i^{(3/2)} +\frac{1}{\pi N} \sum_{j \neq i} \frac{u_j^{(3/2)} -u_i^{(3/2)} }{ u_j^{(0)} -u_i^{(0)}  }- i \pi \frac{ N_f }{2 N} {u_i^{(0)}}^2=0
\end{equation}
which gives a non zero value for  $u_i^{(3/2)}$.
Despite this apparent success, we first note that the procedure
just described seems model dependent and that it can still fail when different field content is considered. Secondly, and more
important, the $R$ charge computed in this way does not match either with the perturbative computation or with
the ${\cal Z}$-extremization result in section \ref{sec:FullZ}. This is a direct evidence that this approach does not allow
to identify the  saddle point that extremizes the partition function.

We now explain how to overcome this problem. We write the integrand of the partition function in a different way: 
the basic idea is that we want to make its
integrand manifestly invariant under the parity symmetry $u_i \rightarrow -u_i$. Indeed, the measure
\begin{equation}
d\Gamma\left(\{u\}\right) \equiv \left( \prod_{i=1}^{N} d u_i \right) \text{Exp}\left(N^2\left( \frac{i \pi}{\lambda N}\sum_{i=1}^{N}u_i^2 +\frac{1}{N^2} \sum_{i\neq j} \log \sinh \left( \pi u_{ij}\right)\right)\right)
\label{eq:measure}
\end{equation}
is invariant under parity, when one also considers that the domain of integration is the whole real axis. For vector-like
theories, the matter contribution is also parity invariant. For chiral theories, this is no longer true, but we can write the
partition function in the following way
\begin{eqnarray} \label{ptfaf2}
\mathcal{Z}& =& \frac{1}{2} \int \prod_{i=1}^{N} d u_i \text{Exp}\left(N^2\left( \frac{i \pi}{\lambda N}\sum_{i=1}^{N}u_i^2 +\frac{1}{N^2} \sum_{i\neq j} \log \sinh \left( \pi u_{ij}\right)\right)\right) \nonumber\\
&\times&\left\{ \text{Exp}\left(\sum_{i} N_f l(1-\Delta_1+i  u_{i})+N_{\widetilde f} l(1-\Delta_2-i  u_{i})\right) \right. \nonumber \\
&\,& \,\, +\left.
\text{Exp}\left(\sum_{i} N_{\widetilde f} l(1-\Delta_2+i  u_{i})+N_f l(1-\Delta_1-i  u_{i})\right)\right\}
\end{eqnarray}
The partition function is exactly the same as the one in (\ref{ptfaf1}) but the symmetry $u_i \rightarrow -u_i$
is now manifest in $\mathcal{Z}$ and in the equations of motion.
The latter are more involved than those derived
from (\ref{ptfaf1}), but they are straightforward to write and solve order by order in $\l$. Moreover, the ansatz (\ref{charge2})
turns out to be consistent at the two-loop level, that is, the solution to the saddle point equations and the partition
function itself share some of the features of the vector-like models.

We computed the solutions both to the saddle point and to the extremization equations and found that the $R$ charge is
\begin{equation}
\Delta_1= \Delta_2 = \frac{1}{2}-\lambda^2 \frac{N_f + N_{\widetilde f}}{4 N}
\label{eq:largendiff}
\end{equation}
which agrees both with the large $N$ limit of (\ref{eq:differentNf}) and, when $N_f = N_{\widetilde f}=N_f$, with
(\ref{eq:sameNf}).

A comment is in order. In the saddle point equations, the matter part of (\ref{ptfaf2}) contributes with a sum
of terms. Each of them is weighted by a factor which reads (we set $N_{\tilde f}=0$ for simplicity, the
generalization is straightforward)
\begin{equation}
\left(1+\exp\left\{ N_f \sum_{i=1}^N \Big[  l\left(1-\Delta_1+iu_i\right) - l\left(1-\Delta_1-iu_i\right) \Big] \right\}\right)^{-1}
\label{eq:exponential}
\end{equation}
Since the sum runs on $N$ terms of order ${\cal O}(1)$, in the large $N$ limit we expect the exponential to be
either $1$, $0$ or divergent when evaluated on the
solution for the $u_i$. If it is not $1$, the large $N$ and the 't Hooft limit do not commute, as we cannot expand
(\ref{eq:exponential}) in powers of $\l$. This would mean that our equations are being solved in an inconsistent way.
If we assume that the eigenvalue distribution is parity invariant, that is, it satisfies $\sum_i \left( u_i^{(j)} \right)^n=0$
for odd $n$ but every $j$, the argument of the exponential in (\ref{eq:exponential}) vanishes. 
Because our solution satisfies this property, we conclude that the large $N$ and
the 't Hooft limit commute, and our result is fully consistent. A similar argument also holds when product gauge groups
are considered.

We also checked, up to the order $\l^{5/2}$ in (\ref{eq:charge3}), that another sufficient condition for the
exponential to be $1$ is that the $R$ charge in (\ref{eq:exponential}) is substituted with its perturbative value
(\ref{eq:largendiff}), without imposing any condition on the eigenvalue distribution.

\subsection{A quiver field theory example: $\widetilde{\mathbb{F}}_0$}
\label{sec:F0}

In this section we apply the result derived above to a more complicated example.
This is a quiver gauge theory which in four dimensions  represents the SCFT living on the world-volume
of a stack of $N$ D$3$ branes probing  a CY$_3$  conical singularity   which has a base
over the Hirzebruch surface or $\mathbb{F}_0$.
\begin{figure}
\begin{center}
\includegraphics[width=5cm]{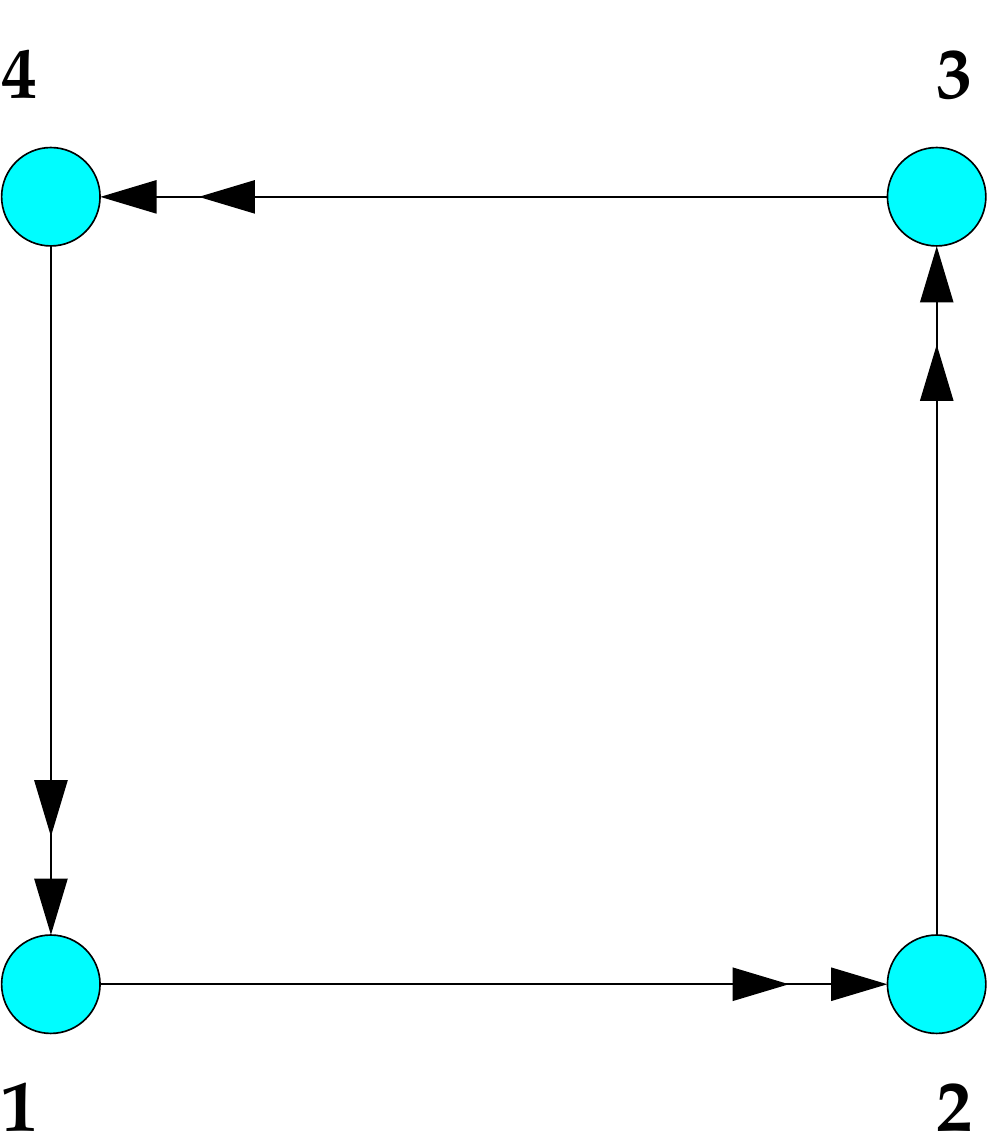}
\end{center}
\caption{Quiver diagram for $\widetilde{\mathbb{F}}_0.$}
\label{FIGFOI}
\end{figure}

The four-dimensional quiver gauge theory consists of a product of four $SU(N)$ gauge groups connected
by chiral bi-fundamental fields as in Figure \ref{FIGFOI}.
The superpotential is 
\begin{equation}
W = \epsilon_{ij} \epsilon_{kl} X_{12}^{(i)}  X_{23}^{(l)}  X_{34}^{(j)}  X_{41}^{(k)} 
\label{eq:WF0}
\end{equation}
In three dimensions
at every gauge group is associated a CS term in the action with level $k_i$.
At large $N$ it is conjectured to describe the  dual field theory
 living on the world-volume of a stack of $N$ M$2$ branes probing  a CY$_4$ conical singularity.
  The geometrical properties of the base of the cone are related to the value of the CS levels. For example in the case of
 $(k_1,k_2,k_3,k_4) = (k,k,-k,-k)$ this SCFT  is conjectured to describe M$2$ branes probing a cone
 over $Q^{2,2,2}/\mathbb{Z}_k$.
 Anyway recently the mismatch in the large $N$ scaling of the free energy among the gravity dual and the field theory side 
placed an obstruction against this conjectured duality.
 
 In the opposite regime $k \gg N \gg 1$ we observe
 a similar problem in the scaling of $\mathcal{F}$. In the case of
 a single gauge group we have been able to overcome the problem by
 restoring the symmetry $u \rightarrow -u$.
Here we follow the same procedure  for the $\widetilde{\mathbb{F}}_0$
 theory, which original partition function is given by
\begin{equation}
{\cal Z} = \int \left( \prod_{I=1}^4 d\Gamma_I \right) f_{mat} \equiv \int \left[ d \Gamma \right] f_{mat}
\label{eq:F0part}
\end{equation}
where we write $d\Gamma_I \equiv d\Gamma(\{u^{(I)}\})$ in equation (\ref{eq:measure}) for brevity.
Here, $f_{mat.}$ is the contribution of the matter fields, that in this case is
\begin{equation}
f_{mat.} = \text{Exp}\left(2 \sum_{I=1}^4 \sum_{i,j} l \left(1-\Delta_{I,I+1} +i(u_{i}^{(I)}-u_{j}^{(I+1)})\right)\right)
\end{equation}
where the $I$s are identified mod $4$ and $\Delta_{I,I+1}$ is the $R$ charge of the fields which connect the
$I$-th and the $(I+1)$-th node.
We then \emph{symmetrize} in analogy with the case with a single group 
obtaining 
\begin{equation}
\mathcal{Z} = \frac{1}{2^4} \int [d \Gamma] g_{mat}
\end{equation}
where the function $g_{mat}$ is 
\begin{equation}
\sum_{Perm.\{ \eta_{K}=\pm 1\} }  \text{Exp}\left(2 \sum_{I=1}^4 \sum_{i,j} l \left(1-\Delta_{I,I+1} +i(\eta_I u_{i}^{(I)}- \eta_{I+1} u_{j}^{(I+1)})\right)\right)
\end{equation}
where the sum is over all the possible $16$ permutations of the set  $\{\eta_K\}$. We restored the full
$\mathbb{Z}_2^4$ reflection symmetry, but up to the two loop order the same procedure also works if one
only makes manifest the $\mathbb{Z}_2$ subgroup which acts by changing the sign of all the four groups
of eigenvalues at once: $\left\{ u_I \right\} \rightarrow - \left\{ u_I \right\}$.

Now by making the ansatz (\ref{charge2}) for the eigenvalues and the $R$ charges we can solve order by order the
saddle point equations and compute the $R$ charges.
Moreover in this case there is  a constraint imposed by the superpotential
\footnote{Note that we should distinguish two different $R$ charges for every couple
of bi-fundamentals connecting two nodes, but the symmetries impose 
$\Delta\left[X_{I,I+1}^{(1)}\right]= \Delta\left[X_{I,I+1}^{(2)}\right]\equiv \Delta_{I,I+1}$.}
\begin{equation}
\sum_{I=1}^{4} \Delta_{I,I+1}=2
\end{equation}
By applying the same technique as above, we get 
\begin{eqnarray}
\Delta_{12}& =& \frac{1}{2} -\frac{\lambda_1^2}{2}-\frac{3\lambda_1\lambda_2}{4}-\frac{\lambda_2^2}{2}
+\frac{\lambda_2 \lambda_3}{4} +\frac{\lambda_3^2}{2}+\frac{\lambda_1 \lambda_4}{4}+\frac{\lambda_3 \lambda_4}{4} +\frac{\lambda_4^2}{2}
+\mathcal{O}(\lambda^4)
\nonumber \\
\Delta_{23}& =& \frac{1}{2} +\frac{\lambda_1^2}{2}+\frac{\lambda_1\lambda_2}{4}-\frac{\lambda_2^2}{2}
-\frac{3\lambda_2 \lambda_3}{4} -\frac{\lambda_3^2}{2}+\frac{\lambda_1 \lambda_4}{4}+\frac{\lambda_3 \lambda_4}{4} +\frac{\lambda_4^2}{2}
+\mathcal{O}(\lambda^4)
\nonumber \\
\Delta_{34}& =& \frac{1}{2} +\frac{\lambda_1^2}{2}+\frac{\lambda_1\lambda_2}{4}+\frac{\lambda_2^2}{2}
+\frac{\lambda_2 \lambda_3}{4} -\frac{\lambda_3^2}{2}+\frac{\lambda_1 \lambda_4}{4}-\frac{3\lambda_3 \lambda_4}{4} -\frac{\lambda_4^2}{2}
+\mathcal{O}(\lambda^4)
\nonumber \\
\Delta_{41}& =& \frac{1}{2} -\frac{\lambda_1^2}{2}+\frac{\lambda_1\lambda_2}{4}+\frac{\lambda_2^2}{2}
+\frac{\lambda_2 \lambda_3}{4} +\frac{\lambda_3^2}{2}-\frac{3\lambda_1 \lambda_4}{4}+\frac{\lambda_3 \lambda_4}{4} -\frac{\lambda_4^2}{2}
+\mathcal{O}(\lambda^4)
\nonumber \\
\label{eq:RF0}
\end{eqnarray}
in agreement with the field theory expectation, with $\l_i \equiv N/k_i$.

If we are not interested in an AdS/CFT example, we can consider the diagram of Figure \ref{FIGFOI}
with an arbitrary number of bi-fundamental fields connecting each pair of nodes.
If  $\a_{I,I+1}$ is the number of fields connecting the $I$-th and the $(I+1)$-th node,
with $\a_{I-1,I}=\a_{I,I+1}$ mod $2$ to avoid parity anomaly,
our procedure gives the large $N$ result
\begin{eqnarray}
\Delta_{12}& =& \frac{1}{2} -\frac{\lambda_1^2 \alpha_{12}}{4}-\frac{3\lambda_1\lambda_2}{4}-\frac{\lambda_2^2 \alpha_{23}}{4}
+\frac{\lambda_2 \lambda_3}{4} +\frac{\lambda_3^2 \alpha_{34}}{4}+\frac{\lambda_1 \lambda_4}{4}+\frac{\lambda_3 \lambda_4}{4} +\frac{\lambda_4^2\alpha_{41}}{4}
+\mathcal{O}(\lambda^4)
\nonumber \\
\Delta_{23}& =& \frac{1}{2} +\frac{\lambda_1^2\alpha_{12}}{4}+\frac{\lambda_1\lambda_2}{4}-\frac{\lambda_2^2 \alpha_{23}}{4}
-\frac{3\lambda_2 \lambda_3}{4} -\frac{\lambda_3^2 \alpha_{34}}{4}+\frac{\lambda_1 \lambda_4}{4}+\frac{\lambda_3 \lambda_4}{4} +\frac{\lambda_4^2\alpha_{41}}{4}
+\mathcal{O}(\lambda^4)
\nonumber \\
\Delta_{34}& =& \frac{1}{2} +\frac{\lambda_1^2\alpha_{12}}{4}+\frac{\lambda_1\lambda_2}{4}+\frac{\lambda_2^2 \alpha_{23}}{4}
+\frac{\lambda_2 \lambda_3}{4} -\frac{\lambda_3^2 \alpha_{34}}{4}+\frac{\lambda_1 \lambda_4}{4}-\frac{3\lambda_3 \lambda_4}{4} -\frac{\lambda_4^2\alpha_{41}}{4}
+\mathcal{O}(\lambda^4)
\nonumber \\
\Delta_{41}& =& \frac{1}{2} -\frac{\lambda_1^2\alpha_{12}}{4}+\frac{\lambda_1\lambda_2}{4}+\frac{\lambda_2^2 \alpha_{23}}{4}
+\frac{\lambda_2 \lambda_3}{4} +\frac{\lambda_3^2 \alpha_{34}}{4}-\frac{3\lambda_1 \lambda_4}{4}+\frac{\lambda_3 \lambda_4}{4} -\frac{\lambda_4^2\alpha_{41}}{4}
+\mathcal{O}(\lambda^4)
\nonumber \\
\end{eqnarray}
which we matched with the two loop diagrammatic  computation.

To conclude this section we observe that our discussion may be relevant at strong coupling, because the 
mismatch observed in \cite{Jafferis:2011zi} with the expected AdS/CFT results should be ascribed
to some problem in the identification of the saddle point, and the symmetrization of the integrand of $\mathcal{Z}$ 
may help in understanding the large $N$ scaling.

\section{Lagrange multiplier} \label{sec3}

It is interesting to see whether the saddle point approximation can correctly describe the RG flow of the $R$ symmetry.
As in \cite{Amariti:2011xp} we here apply the technique
of \cite{Kutasov:2003ux,Kutasov:2004xu} to answer this point.

For $\mathbb{F}_0$, we consider a modified version of the partition function of section \ref{sec:F0}
\begin{equation}
\widetilde {\cal Z} = \int [d \Gamma] \widetilde f_{mat.}
\end{equation}
where $[d \Gamma]$ for this case is defined in section \ref{sec:F0} and $\widetilde f_{mat}$ is
\begin{eqnarray}
\widetilde f_{mat}&=& \text{Exp}\left(2 \sum_{I=1}^4 \sum_{i,j} l \left(1\!-\!\Delta_{I,I+1} \!+\! i(u_{i}^{(I)}\!-\!u_{j}^{(I+1)})\right)\! + \! m(\l_1) \left( \sum_{I=1}^4 \Delta_{I,I+1} - 2 \right) \right) \nonumber \\
m(\l_1) &=& m_0 + m_2 \l_1^2
\end{eqnarray}
In the following discussion, $m(\l_1)$ will play the role of a Lagrange multiplier which enforces the marginality constraint
from the superpotential (\ref{eq:WF0}). A couple of comments are in order. We added only one Lagrange multiplier.
As long as the number of bi-fundamental fields connecting two different nodes is the same this is perfectly consistent.
Had we chosen a different number of fields connecting the different nodes, the renormalization group equations would
have not preserved the whole symmetry of the superpotential (\ref{eq:WF0}) away from the fixed point, and we
would have to add more multipliers.
 Secondly, we chose the Lagrange multiplier to be a function of $\l_1$ only. This is
consistent because we are assuming that all the $\l$'s are small. Thus, we will write $\l_i=\l_1 \frac{\l_i}{\l_1}$
and we will expand our saddle point equations in powers of $\l_1$. As long as $\l_i \ll 1$, $i=1,\ldots,4$ this  gives 
the right result.

We symmetrize the first term in $\widetilde f_{mat}$ as discussed in the previous sections, and derive the saddle point equations
which are not affected by the Lagrange multiplier. Then, we write the four extremization equations which simply reads
\begin{equation}
\frac{\partial}{\partial\, \Delta_{I,I+1}}\, \log{\widetilde f_{mat}} = 0 \qquad I=1,\ldots,4
\label{eq:LEF0}
\end{equation}
and compute the solution as a function of $m(\l_1)$. At the fixed point, after we substituted back into the partition
function the solution to (\ref{eq:LEF0}), the equation
\begin{equation}
\frac{\partial}{\partial\, m(\l_1)}\, \log{\widetilde f_{mat}} = 0
\end{equation}
also holds. It is the latter equation which enforces the marginality of the superpotential; once it is imposed, it allows us
to find the fixed point superpotential coupling $h$, which is related to $m$, as a function of the fixed point 't Hooft couplings
$\l$'s.

We do not discuss the solution in details, as the procedure used and the solution itself are very similar to those in section
\ref{sec:F0}. Thus, we only present the results, mentioning that the equations set $m_0=0$ and
\begin{eqnarray}
\Delta_{12}& =& \frac{1}{2} - \lambda_1^2 - \l_1 \lambda_2 - \lambda_2^2 - \frac{m_2 \l_1^2}{\pi^2}
+\mathcal{O}(\lambda^4)
\nonumber \\
\Delta_{23}& =& \frac{1}{2} - \lambda_2^2 - \lambda_2 \lambda_3 - \lambda_3^2 - \frac{m_2 \l_1^2}{\pi^2}
+\mathcal{O}(\lambda^4)
\nonumber \\
\Delta_{34}& =& \frac{1}{2} - \lambda_3^2 - \lambda_3 \lambda_4 - \lambda_4^2 - \frac{m_2 \l_1^2}{\pi^2}
+\mathcal{O}(\lambda^4)
\nonumber \\
\Delta_{41}& =& \frac{1}{2} - \lambda_1^2 - \lambda_1 \lambda_4 - \lambda_4^2 - \frac{m_2 \l_1^2}{\pi^2}
+\mathcal{O}(\lambda^4)
\nonumber \\
\end{eqnarray}

By comparison with the perturbative result (\ref{eq:RF0}), we obtain
\begin{equation}
m_2 \l_1^2 = \frac{|h|^2 N^2}{16}
\end{equation}
which is the expected relation from the discussion in \cite{Amariti:2011xp}.

\section{Discussion}
\label{sec:end}

In this paper we have studied the large $N$ behavior of the free energy
at the perturbative level for both vector and chiral like gauge theories.
We carried out the analysis by solving the saddle point equations
for the eigenvalues with an appropriate ansatz.
We observed that in the non-chiral case this procedure reproduces the 
two loop calculations, while in the case of theories with a chiral matter content 
the saddle point equations cannot be solved order by order in the
't Hoof coupling.
We have shown that this problem can be overcome by restoring the Weyl symmetry 
over the Cartan of every gauge group on the saddle point equation.
This symmetrization acts on the integrand but it does not modify the free energy, 
which is indeed integrated over the Cartan  subgroup.
We have shown that after this transformation, a convenient ansatz correctly solves the equations
at least at the lowest order in the 't Hooft coupling, and by extremizing the free energy around the
saddle point the two loop field theory results have been reproduced.

It would be important to go beyond the two loop approximation and observe whether the procedure that 
we have worked out in this paper does still apply. There are two interesting checks.
The first consists of matching the $\lambda^4$ order which can be obtained with our procedure with the direct 
computation of the partition function at large $k$ but finite $N$, where the large $N$ limit can be
safely taken after the integration over the variables $u$. 
This is just a consistency check for the computation of the partition function.
A more complete check consists of matching with the four loop
perturbation theory.

As already observed in the introduction this paper does not address the problem of the large $N$ 
behavior of chiral like gauge theories at strong coupling.
In many cases the expected result can be computed from the AdS/CFT correspondence.
Indeed if a chiral gauge theory describes the motion of M$2$ branes  probing 
a toric CY$_4$ cone over a Sasaki Einstein manifold Y, then the free energy is related to the volume of Y.
Recently in \cite{Gulotta:2011aa} it has been observed that 
the computation of the volumes matches with the counting  of 
the number of gauge invariant operators
with a given $R$ and monopole charge.

 This procedure was also applied to
the $M^{1,1,1}/\mathbb{Z}_k$ case, and it was observed that this counting
matches with the geometrical computation of the volumes
as a function of the trial $R$ charges. \footnote{The conjecture have been even  tested for the  $Q^{2,2,2}/\mathbb{Z}_k$
model, but only after imposing the exact $R$ charges.}  
 Anyway this counting does not match with the eigenvalues distribution obtained from the
saddle point equations of the free energy in the chiral cases. 
It would be worth to obtain the $N^{3/2}$ scaling of the free energy and the matching 
among the field theory and the supergravity computation from a purely field theoretical extremization of 
 $\mathcal{F}$  and see whether the conjecture above holds. 
We think that our procedure can give some 
hints to study this problem.
Indeed our main result consists in rewriting the partition function such that some symmetries
become explicit in the saddle point equations. These restored 
symmetries then allow to maintain
the same ansatz conjectured for vector like theories also in the chiral case.
It is possible 
 that also at strong coupling a similar procedure would allow for a consistent solution
of the saddle point equations in terms of the common ansatz $u_i = \sqrt N x_i+i  y_i$
 , which leads to the expected result ${\cal F} \propto N^{3/2}$.

\section*{Acknowledgments}

We are grateful to P.~Agarwal, C.~Closset, K.~Intriligator, C.~Klare and A.~Zaffaroni for discussions.
A.A. is supported by UCSD grant DOE-FG03-97ER40546.  The work of M.S. is supported in part by the FWO - Vlaanderen,
Project No. G.0651.11, and in part by the Federal Office for Scientific,
Technical and Cultural Affairs through the ``Interuniversity Attraction
Poles Programme -- Belgian Science Policy'' P6/11-P.

\appendix

\section{Monopoles} \label{appmax}

Let us consider the partition function for a ${\cal N}=2$ Chern-Simons theory coupled to $N_f$ fundamental fields
\begin{equation}
{\cal Z} = \int\!\! \prod_i du_i \,\, e^{i k \pi \sum_i u_i^2} \, e^{-\Delta_m \sum_i u_i} \, \prod_{i<j} \sinh^2\left( \pi u_{ij} \right) \, e^{N_f \sum_i l\left(1-\Delta+i u_i \right)}
\label{eq:partapp}
\end{equation}
We write the argument of the CS and monopole contributions as
\begin{equation}
ik\pi \left( u_i^2 + i \frac{\Delta_m u_i}{k\pi} \right) = ik \pi \left( u_i + i \frac{\Delta_m}{2\pi k} \right)^2 + i \frac{\Delta_m^2}{4\pi k}
\label{eq:expapp}
\end{equation}
We shift the integration variables as
\begin{equation}
u_i = u_i^\prime - i \frac{\Delta_m}{2\pi k}
\label{eq:shiftapp}
\end{equation}
and we substitute (\ref{eq:expapp}) and (\ref{eq:shiftapp}) into (\ref{eq:partapp})
\begin{equation}
{\cal Z} = e^{ i \frac{\Delta_m^2}{4\pi k}} \int\!\! \prod_i du_i \,\, e^{i k \pi \sum_i u_i^2} \, \prod_{i<j} \sinh^2\left( \pi u_{ij} \right) \, e^{N_f \sum_i l\left(1 - \Delta + \frac{\Delta_m}{2\pi k} + i u_i\right)}
\label{eq:Zmonopole}
\end{equation}

The last formula makes two things manifest. First, the charges only appear in the combination
$\Delta-\frac{\Delta_m}{2\pi k}$. When the gauge group is $U(N)$, we can impose $\sum_{SU(N)} u_i=0$ so that
only the $U(1)$ weight in (\ref{eq:shiftapp}) is shifted. Then, $\Delta_m$ in (\ref{eq:Zmonopole})
only appears in the $U(1)$ contribution. This corresponds to the fact that gauging a $U(1)$ symmetry one
introduces a gauge field which couples to $j_{matter} + k j_{top}$. Secondly, in the perturbative regime $k \gg 1$, the monopole
contribution is suppressed by a factor $1/k$ with respect to the contribution from the $R$ charge $\Delta$.
Thus, at least up to the two loop order, the monopole charge vanishes, and the difference between the
$SU(N)$ and the $U(N)$ theories is only due to the different values of the Casimir operators. In
the large $N$ limit this difference is subleading in $N$ and the two models coincide,
as expected from standard perturbation theory.

\bibliographystyle{JHEP}
\bibliography{Large}

\end{document}